\documentclass[conference]{IEEEtran}

\ifCLASSINFOpdf
   \usepackage[pdftex]{graphicx}
\else
\fi
\usepackage{flushend}
\begin{document}
%
\title{Building Block Components to Control a Data Rate  in the Apache Hadoop Compute Platform}

\author{\IEEEauthorblockN{Tien Van Do\IEEEauthorrefmark{1}, Binh T. Vu\IEEEauthorrefmark{1}, Nam H. Do\IEEEauthorrefmark{1},
L\'or\'ant Farkas\IEEEauthorrefmark{4}, Csaba Rotter\IEEEauthorrefmark{4}, Tam\'as Tarj\'anyi\IEEEauthorrefmark{4}}
\IEEEauthorblockA{\IEEEauthorrefmark{1}Analysis, Design and Development of ICT systems (AddICT) Laboratory\\
Budapest University of
Technology and Economics\\
Magyar tud\'osok k\"or\'ut 2, Budapest, Hungary
\\
Email: do@hit.bme.hu
}
\IEEEauthorblockA{\IEEEauthorrefmark{4}Nokia  Networks\\
K\"oztelek utca 6, Budapest, Hungary}
}

%



\maketitle

\begin{abstract}
Resource management is one of the most indispensable components of cluster-level infrastructure layers.
Users of such systems should be able to specify their job requirements as a configuration parameter (CPU, RAM, disk I/O, network I/O) and have the scheduler translate those into an appropriate reservation and allocation of resources.
YARN is an emerging resource management in the Hadoop ecosystem, which supports only RAM and CPU reservation  at present.

In this paper, we propose a solution that takes into account the operation of
the Hadoop Distributed File System
 to control the data rate of applications
in the framework of a Hadoop compute platform. We utilize the property that
a data pipe between a container and a DataNode consists of  a disk I/O  subpipe and a TCP/IP subpipe.
 We have implemented building block software components
to  control the data rate of data pipes between containers and DataNodes and provide a proof-of-concept with
measurement results.
\end{abstract}





%
\IEEEpeerreviewmaketitle

\section{Introduction}




Heterogeneous compute clusters
can be easily established using physical
machines incorporating memory, disks and powerful CPUs
to offer Information and Communications Technology (ICT) services.
Hadoop \cite{Hadoop,White2012,Shvac2010,Vavi2013} is a software framework that has been developed to satisfy the need of processing data
in the scale of petabytes/day (i.e., big data \cite{ThusooSJSCZALM10,Zi}) with the use of resources offered by compute clusters.

The design of  Hadoop considers several factors such as reliability,
scalability, programming model diversity, flexible resource model etc~\cite{Vavi2013}.
The popularity of Hadoop is mainly due to a design decision that
allows parallel and distributed computing meanwhile it
hides a complexity
from users \cite{White2012,Shvac2010,Polato20141}.
Originally, Hadoop consisted of a distributed file system and a MapReduce processing framework.
Later on, the need for supporting different processing paradigms was recognized and YARN was introduced~\cite{Vavi2013,Yarn}.

Application layer software (customer experience management, operation support system, customer care are typical examples in telecommunication environments) use the services of the infrastructure layer (e.g. Hadoop) and they together reserve resources through the platform (OS, kernel, firmware, etc.), like CPU, RAM disk I/O and network I/O. Scheduler translates those reservations into an appropriate allocation of resources~\cite{2013Barroso}.

Job scheduling is an additional task of resource management~\cite{Vavi2013,Jenning2014,Suresh201470,Hindman2011,Thain:2005}, which is one of the most indispensable components of cluster-level
infrastructure layers.
YARN~\cite{Yarn} is a distributed resource management system for resource allocation in compute clusters~\cite{White2012,Vavi2013} and
job scheduling, in the particular case of Hadoop MapReduce.
It is recognized that job scheduling is a challenging issue  because
various factors (quality of service, performance,  etc)
should be taken into consideration to improve the
degree of satisfaction of users.

In mobile network environments network equipment vendors are increasingly facing the challenge that their solutions and products need to be deployed in a so called white box scenario, where they run on the same physical infrastructure as applications of the mobile operator, and even more, they share some of the cluster level infrastructure (e.g. shared Hadoop cluster).
In such a scenario, a typical Big Data application may consist of multiple jobs
that are executed a distributed manner (up to several thousands machines).
Some customers may require a data rate guarantee
because jobs should be finished by a certain deadline.
Therefore, the provision of the quality of service regarding a data rate guarantee may play a key factor to attract customers.
However,   YARN (up to version 2.5.1~\cite{Yarn}) only supports the reservation of memory and CPU in compute clusters at present.

In this paper, we exploit the special feature of the
Hadoop Distributed File System --HDFS  (which is the part of Hadoop~\cite{Hadoop})
and  the capability
of Linux Traffic Control --LTC  (which was developed under Linux kernels 2.2 and 2.4 and
now is incorporated in the newest Linux kernels as a module)  subsystem to control the data pipes
of containers to HDFS DataNodes in YARN. We
 propose building block software components
that can be integrated into YARN to control the data throughput of applications.
We use ZooKeeper~\cite{ZooKeeper} to maintain persistent information to control the throughput of data pipes.


The rest of this paper is organized as follows. In Section~\ref{sec:bg}, some technical backgrounds on Hadoop, HDFS and resource management are presented.
In Section~\ref{sec:pro}, a proposal is described. In Section~\ref{sec:proof}, a proof-of-concept is illustrated with measurement results.
Finally, Section~\ref{sec:con} concludes our paper.



\section{Technical Background}
\label{sec:bg}
In this Section, we provide a short summary of features and properties we use to construct
our proposed solution.

\begin{figure}[hbt]
		\centering
		\includegraphics[scale=0.35]{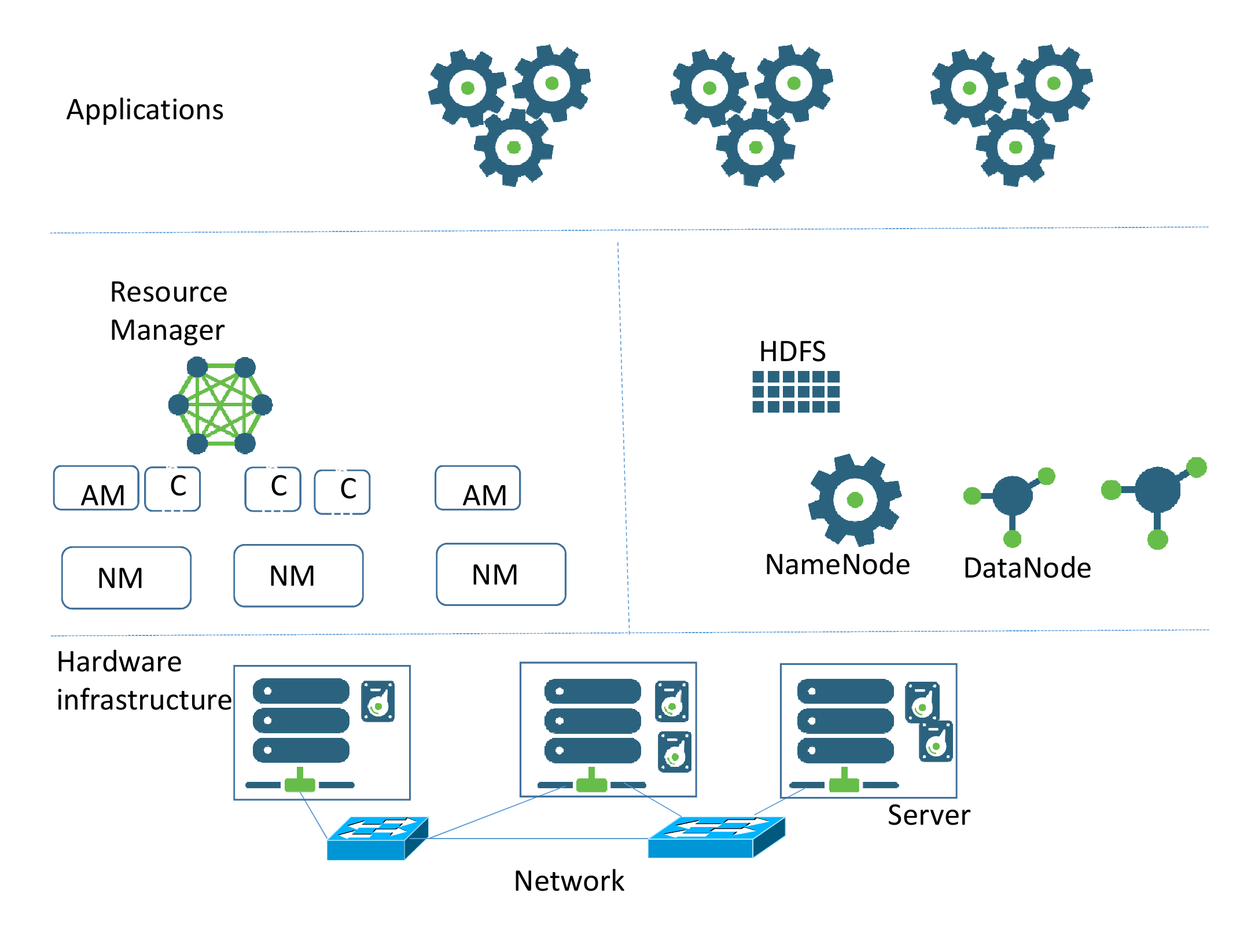}
		\caption{Main components in a Hadoop Compute Platform}
		 \label{fig:layer}
\end{figure}

A Hadoop (version 2.0 or higher) compute cluster normally consists of four main groups
(one hardware group and three software groups) that are illustrated in
Figure~\ref{fig:layer}:
\begin{itemize}
\item The hardware infrastructure  includes
a platform of machines/servers with CPUs and disks, and a network
that connects the machines.
The hardware infrastructure and the operating systems (running directly on physical or virtual machines) provides resources for the Hadoop system and applications.

\item The Hadoop Distributed File System (HDFS) with functional entities (NameNode and DataNodes) is a distributed file system that runs
on physical or virtual machines. It stores large data sets (files of gigabytes to terabytes) and provides
streaming data access to clients and applications.
\item The resource management group with functional entities (Resource Manager, NodeManagers, ApplicationMasters) is responsible to process the resource requirement of applications, and
decides where (which physical machines) a specific application should run based on the knowledge of the hardware
infrastructure and the locations of  blocks in the HDFS  storage.
\item Applications analyze Big Data and do some computations on Big Data. One type of applications use MapReduce, that is
a programming model for data processing~\cite{Shvac2010,White2012} and  imposes a typical workload on top of HDFS consisting of 3 phases associated with the 3 phases of the processing paradigm, map, shuffle and reduce.
Another type of applications of the HDFS is HBase, which is a  distributed,  column-oriented  database
to support real-time read/write random access to
very large datasets~\cite{White2012,Shvac2010}. A systematic survey of application types and their characteristic workload on HDFS is beyond the scope of this paper.
\end{itemize}


\subsection{Operation of HDFS}

In the Hadoop, NameNode and Datanodes are functional components that realize
the Hadoop Distributed File System to store files and retrieve blocks of data~\cite{White2012,Shvac2010}.
If used as a file system, then files are splitted into blocks and stored in Datanodes. To ensure the reliable service against failures,
the replication mechanism  may be applied to allow the placement of the same
blocks in different Datanodes.
The NameNode is responsible for storing the filesystem tree, the metadata of all the files
and directories in the HDFS file system.
Information about the locations of the blocks of a specific file is also maintaned by the NameNode.

\begin{figure}[hbt]
		\centering
		\includegraphics[scale=0.35]{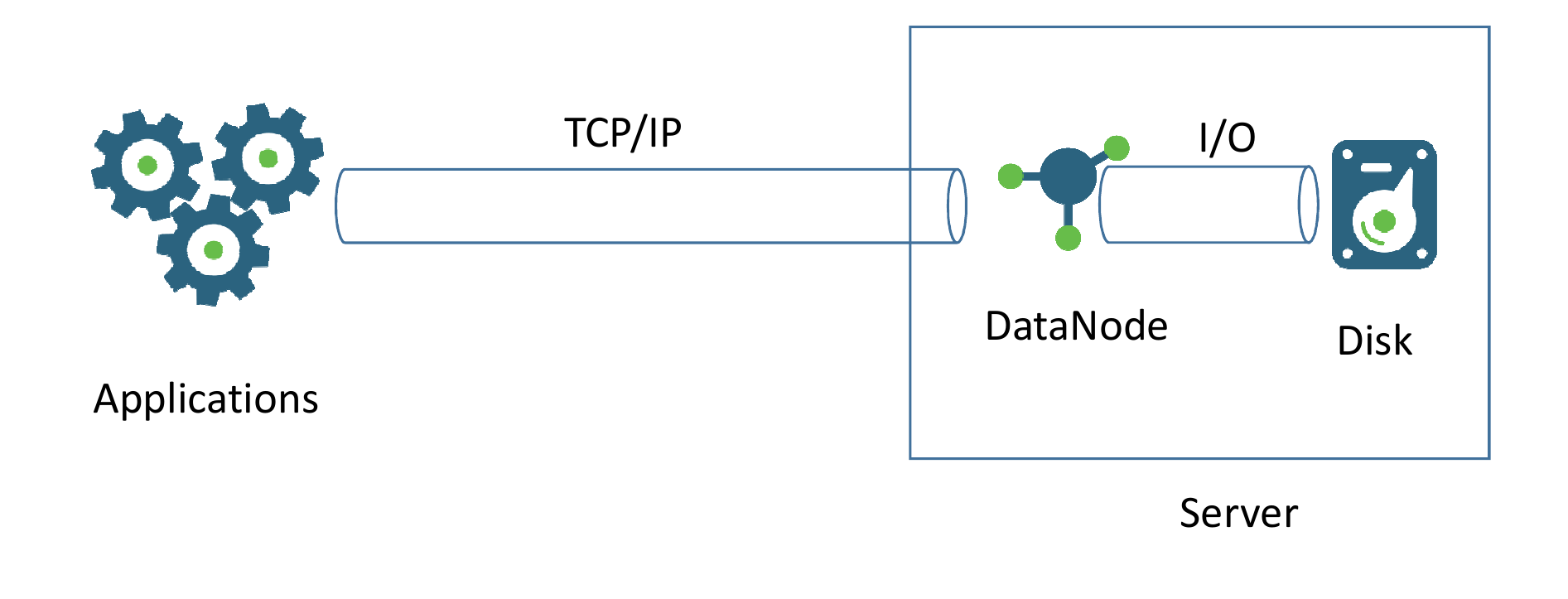}
		\caption{Data pipes between an application and a DataNode}
		 \label{fig:pipe}
\end{figure}

To access/read a specific file, an HDFS client initiates a request to the NameNode to enquire
about the list of Datanodes that stores replicas of the blocks of the file. Then, the
HDFS client chooses a Datanode that  stream data blocks to the client (see Figure~\ref{fig:pipe}).
In HDFS all communications and data transfers are performed using the TCP/IP protocol stack.

It is worth emphasizing that the streaming  data of a specific file block
is conveyed through two pipes (as illustrated
in Figure~\ref{fig:pipe}): a TCP/IP pipe (through either  a network or the loopback interface of a Datanode's machine) between
an application and a DataNode,  and a disk I/O pipe between a DataNode and a certain disk.

\subsection{Resource Management}

YARN decouples the programming
model from the resource management infrastructure~\cite{Vavi2013}.
In the YARN architecture there are several important entities: Resource Manager (RM), Node Managers (NM),
Application Masters (AM).
There is a special term \textquoteleft\textquoteleft container\textquoteright\textquoteright $\;$ that
is the collection of resources (CPU and memory) centrally assigned by the RM.
In YARN, negotiations  regarding resources are performed
between  a client, its ApplicationMaster and RM,
and decisions are taken by RM.  

However, the resource usage related to HDFS storage is not covered by YARN, which may cause performance problems
because certain types of applications such as MapReduce have HDFS-intensive resource consumption.
Furthermore, the identification of HDFS data pipes is hidden from other resource management functions,
which causes a challenge for resource management.
We describe our approach in Section \ref{sec:pro} to handle these problems.




\begin{figure*}[hbt]
		\begin{center}
		\includegraphics[scale=0.4]{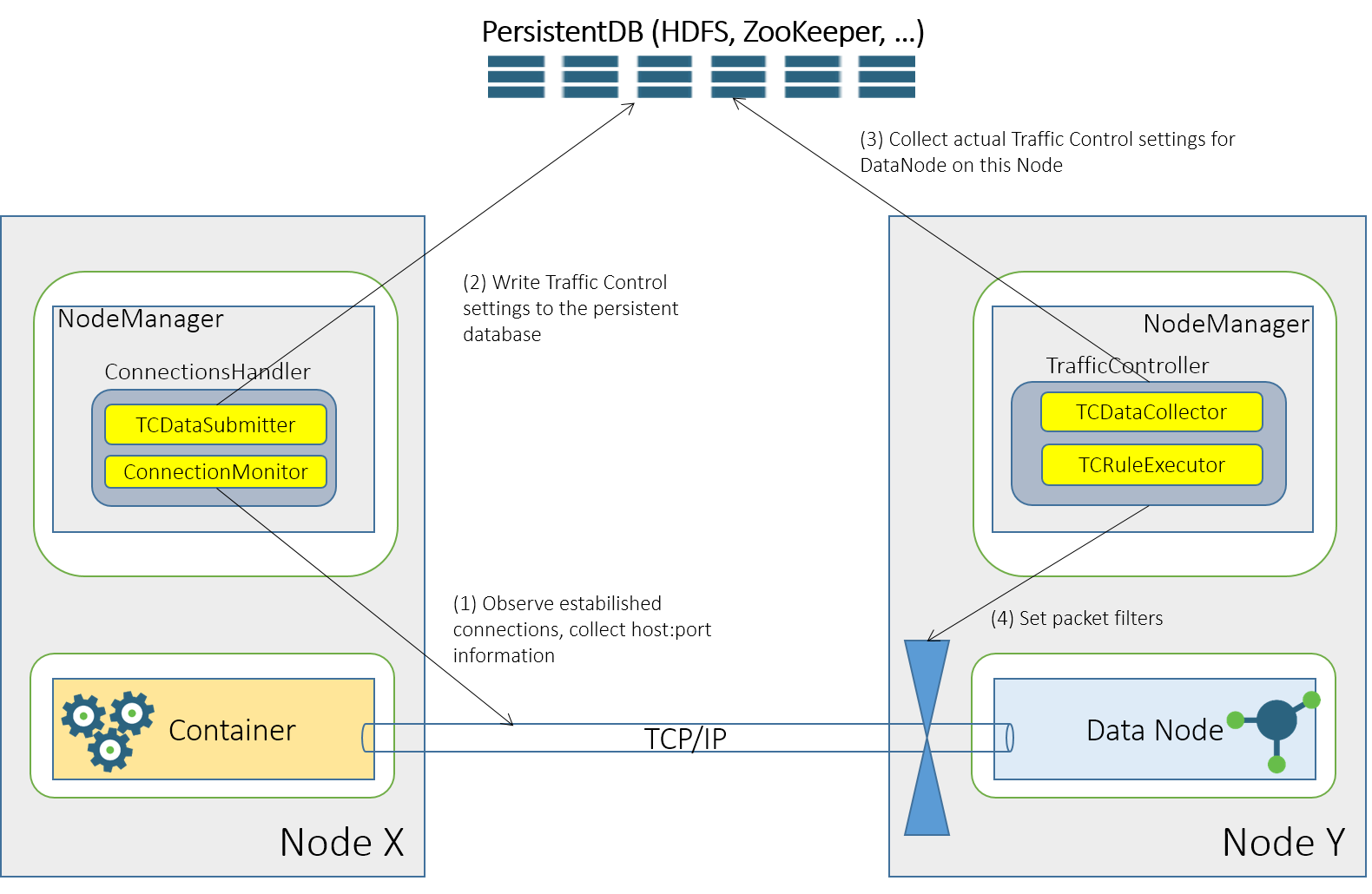}
		\caption{Control the rate of pipes} 		 \label{fig:impl}
		\end{center}
\end{figure*}

\section{A Proposed Solution}
\label{sec:pro}
In this Section, we propose
 a solution that allows service providers to control
the data rate (customer's QoS requirement) of applications from an HDFS storage in Hadoop compute clusters.

Quality of Service (QoS) is defined by Recommendation ITU-T G.1000~\cite{ITUG1000}
 as  the collective effect of service performances that characterize the
degree of satisfaction of a user. There are several QoS criteria (speed, accuracy, availability, reliability,
security, simplicity and flexibility)~\cite{Richter88} that serves as the base for setting QoS
parameters and performance objectives.
Furthermore, there are four viewpoints~\cite{ITUG1000} of QoS from the perspective of customers and service providers:
customer's QoS requirements, QoS offered by a provider, QoS achieved by a provider, QoS perceived by a customer.
It is worth mentioning that mechanisms (rules, procedures, policies) should be deployed in the infrastructure of service providers to
provision QoS for customers.

\begin{figure*}[hbt]
		\centering
		\includegraphics[scale=0.45]{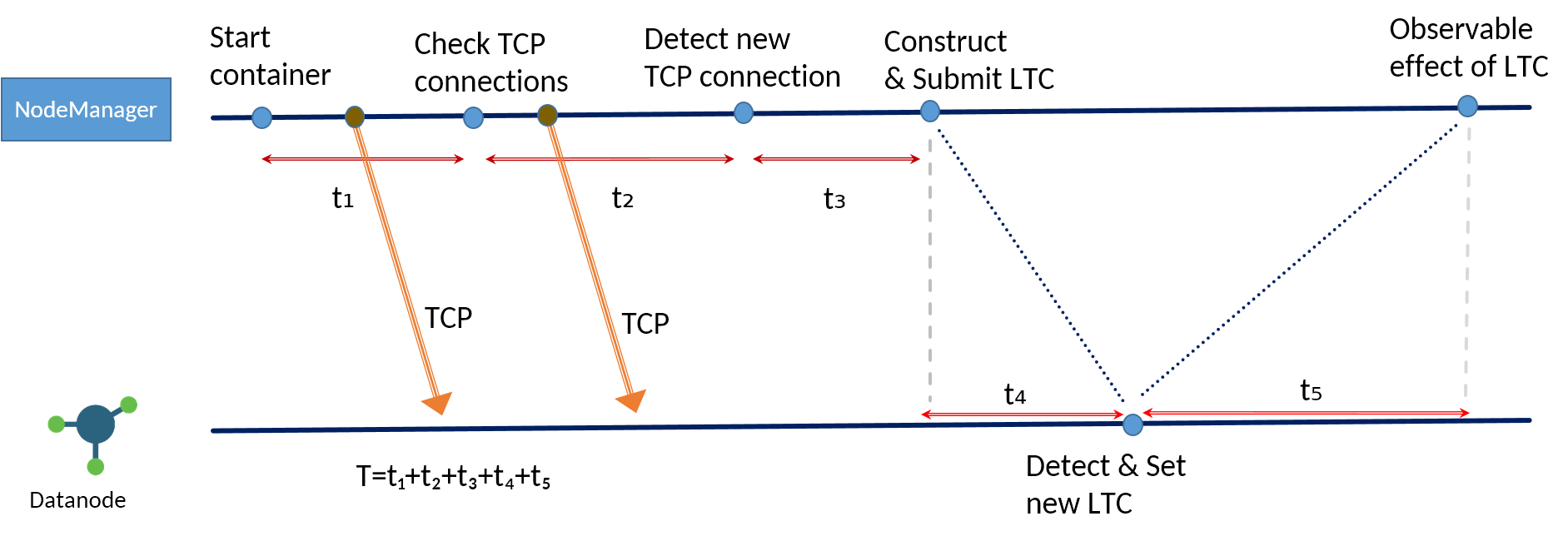}
		\caption{The time line of the flow execution}
		 \label{fig:time_flow}
\end{figure*}

Today YARN is not supporting yet metrics like delay/latency of the query execution, which would be very useful to specify/enforce from the viewpoint of an end user or application. YARN is dealing with amount of CPU and RAM. The translation from the language of CPU cycles and RAM gigabytes to metrics like query execution type is far from trivial and deep application insight/fingerprinting is necessary, which is subject for future work.
Following the general principles regarding the provision of QoS from the viewpoint of service provider,
   mechanisms to control  the data rate of applications should include
\begin{itemize}
\item the specification of requirements of users (applications),
 \item
Information about a compute cluster, i.e.,
\begin{itemize}
\item  the maximum capacity of the resource of a cluster
(i.e., the maximum capacity of disk I/O), 
\item the network topology and the network capacity (i.e., the maximum capacity of network I/O between machines)
of a cluster, 
\item  the amount of resource occupied by
containers in a cluster, and
the identification of pipes between applications and DataNodes, and
amongst DataNodes,
\end{itemize}

\item resource management policy (i.e., a strategy to allocate resource) and decision procedures (admission control and policing) performed by RM.
\end{itemize}

Within the YARN framework, a client submit a job to the RM.
The submission of a job contains the resource requirements for the container
 that will host the ApplicationMaster of the client~\cite{White2012,Vavi2013}.
ApplicationMaster is responsible to  request a set of containers to run its tasks on.
An instance of {\tt Resource} class conveys the resource type requirements of containers. Therefore, to support a new type of resource,
{\tt Resource} class should be extended to contain the requirements of a new resource type (e.g., the IOPS,
 the reads per second, the writes per seconds, throughput).

Upon the arrival of requests for containers, RM should perform
an admission control procedure to check whether current available resources
are sufficient for  the requested containers.
The decision by the admission control procedure is based on the information about the
capacity of the cluster, the amount of resources occupied by the allocated containers in the cluster and
the resource requirement of containers that are being requested by
the ApplicationMaster of a specific client.

If the admission control allows the allocation of a container, the ApplicationMaster sets up
the ContainerLaunchContext and communicate with the ContainerManager to start its allocated container.
The ApplicationMaster also monitors the status of the allocated containers.
If a task running on a container finished,
the
ApplicationMaster will get updates of completed containers.

A policing function is responsible to keep and guarantee the required resource for
the allocated containers. To control the data rate, the
operation of the Hadoop Distributed File System (i.e., how files and blocks are streamed by DataNodes to applications)
is  taken into consideration in what follows.

Since the streaming  data of a specific file block
is conveyed through two pipes as illustrated
in Figure~\ref{fig:pipe},
\begin{itemize}
\item
the enforcement of the I/O usage of containers and HDFS could be done
   at the HDFS DataNodes in the machine level,
\item
   all the I/O activities of tasks depends on the TCP connections handled by
   HDFS datanodes,
\item
 the throughput of TCP connections can be controlled by Linux Traffic Control (LTC) --see Figure \ref{fig:impl}.
\end{itemize}

In Linux, queues\footnote{http://lartc.org/howto/lartc.qdisc.html} can be setup to manage the bandwidth of TCP/IP pipes.
That is, filters  can be specified to classify traffic based on the source address, the destination address, and the port numbers
of TCP sessions, and/or u32\footnote{a match on any part of a packet} condition. Then, different algorithms (e.g., Token Bucket
Filter, Stochastical Fairness Queueing, Random Early Detection) can be used to control the rate of TCP/IP sessions~\cite{Hubert}.









Due to the decoupling of functionality, the opening of data pipes (Figure~\ref{fig:pipe}) for the usage of
the HDFS storage is not explicitly covered by the resource negotiation process.
However, applications like MapReduce intensively access the blocks of big files stored in HDFS.
Furthermore, the identification of data pipes is hidden from other resource management functions
and can not be revealed at the beginning (e.g., which DataNode is to
be contacted for a certain data block by a specific MapTask), which causes a challenge for resource management.
To configure LTC and control a TCP pipe, the information about the existence of pipes must be obtained. For this purpose,
we either take the creation of pipes
to the negotiation process or implement a monitor function that senses the setup of TCP pipes between containers and DataNodes.
Figure~{\ref{fig:time_flow}} illustrates the time line of the flow execution for the latter alternative
with the following durations from the aspect of controlling pipes:
\begin{itemize}
\item $t_1$ is the duration between starting the container and begin of checking TCP connections.
Note that TCP connections may be existing in this time interval.
\item $t_2$ is the duration needed to  detect TCP connections.
\item $t_3$ is the duration to  construct and submit new LTC settings. 
\item $t_4$ is the duration needed to configure LTC for a DataNode.
\item $t_5$ is the latency between the start of control and observable effect.
\end{itemize}
It will be shown in Section~\ref{sec:proof} that the delay ($T=t_1+t_2+t_3+t_4+t_5$) are acceptable for certain cases (especially when
big HDFS blocks are streamed).
\begin{figure}[hbt]
		\centering
		\includegraphics[scale=0.40]{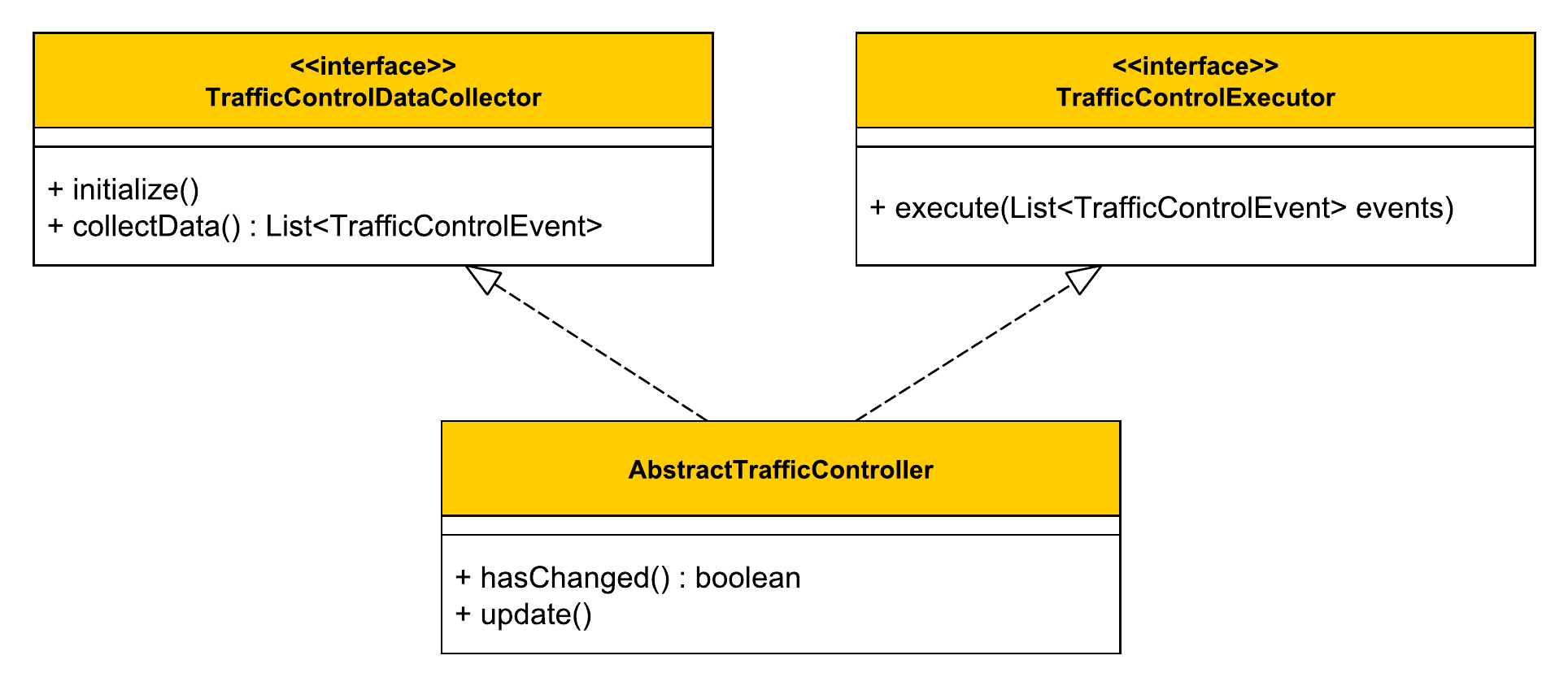}
		\caption{Interface for DataNodes}
		 \label{fig:idata}
\end{figure}
\begin{figure}[hbt]
		\centering
		\includegraphics[scale=0.40]{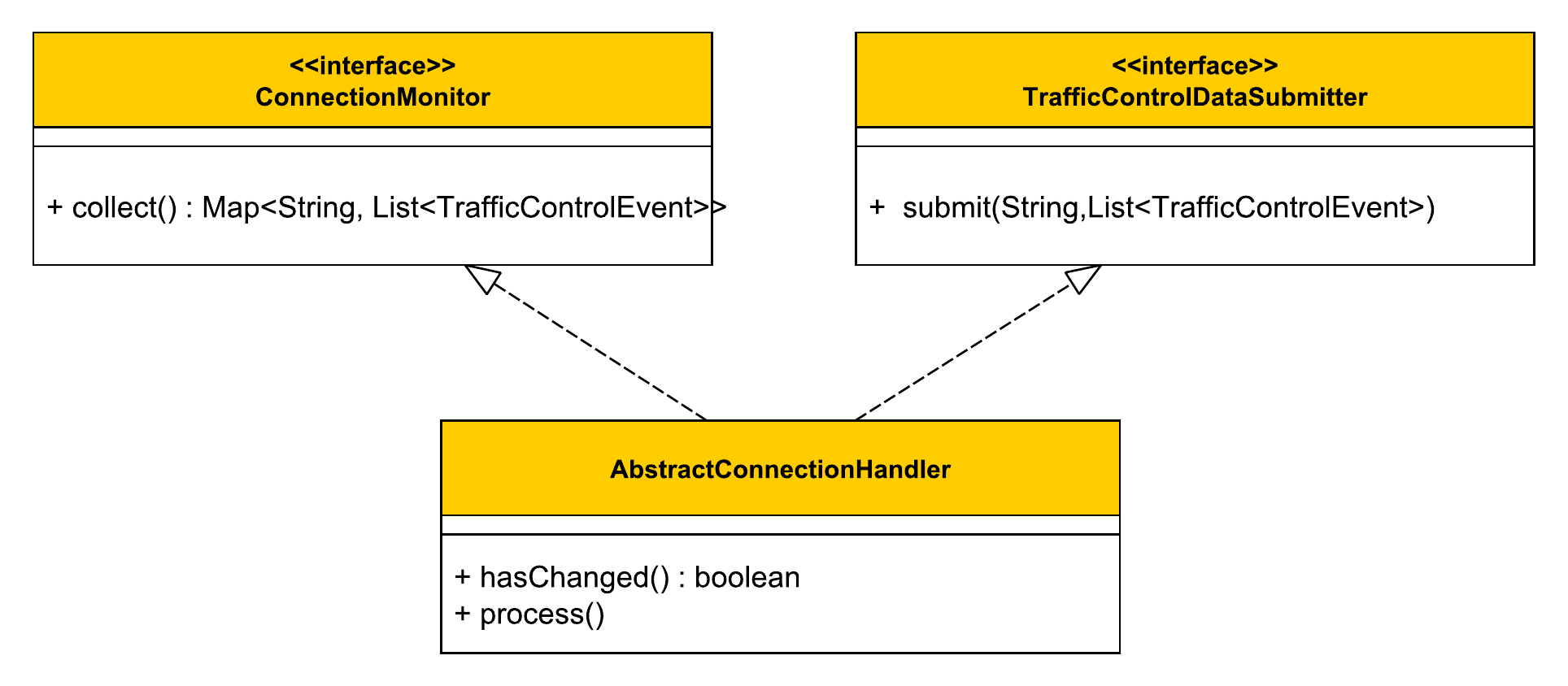}
		\caption{Interface in machines with a NodeManager}
		 \label{fig:idata2}
\end{figure}

\begin{figure*}[hbt]
		\centering
		\includegraphics[scale=0.45]{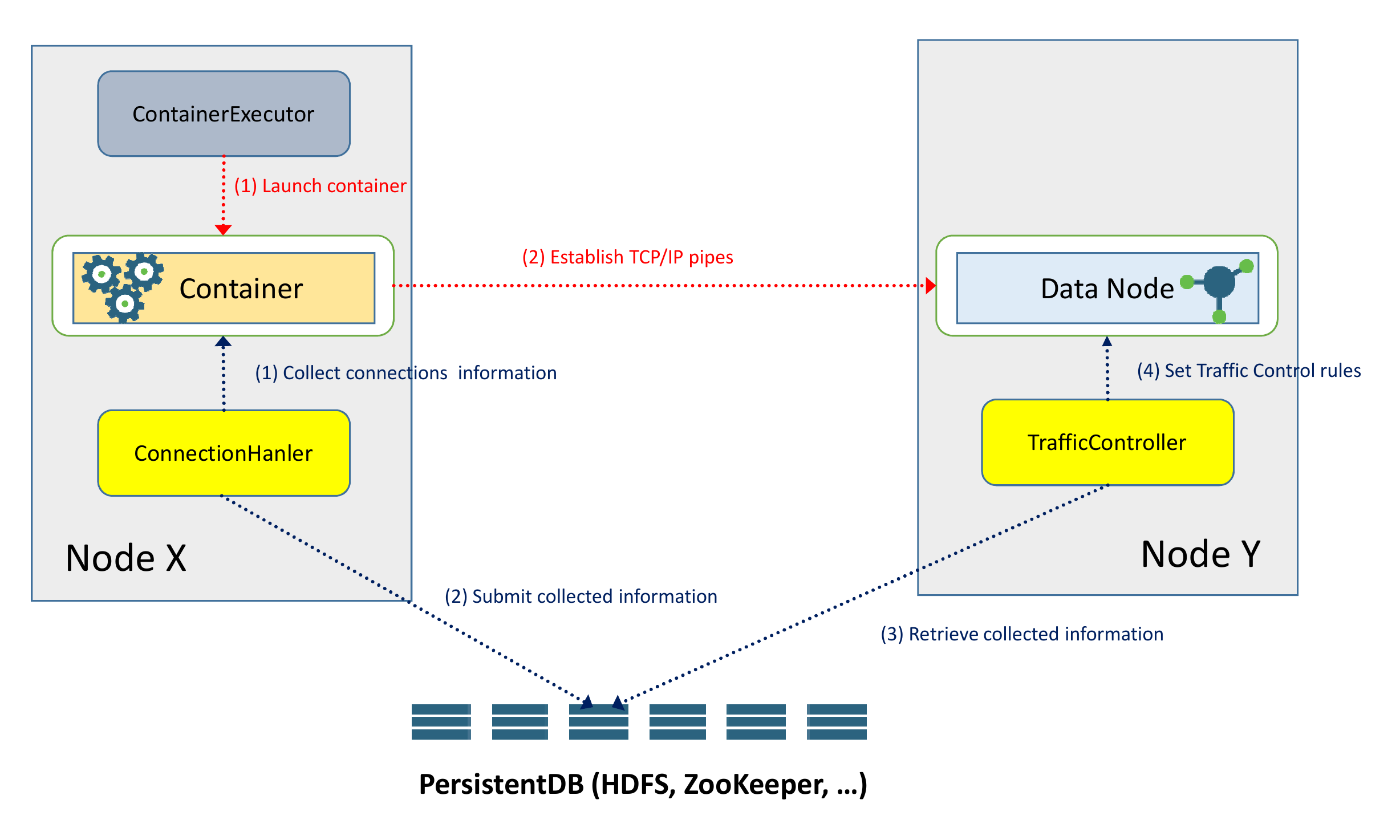}
		\caption{The flow of executions when TCP connections are detected}
		 \label{fig:flow}
\end{figure*}

\subsection{Interfaces}
It is worth mentioning that there are a number of alternatives to implement mechanisms and procedures. Therefore,
our approach is  to define clear interfaces between functions and mechanisms and implement building-block functions (see illustration
in Figures~\ref{fig:idata} and~\ref{fig:idata2} where interfaces and functions for the exchange of information and setting LTC are shown):
\begin{itemize}
\item
{\tt ConnectionMonitor} maintains the information of connections between container nodes and DataNodes.

\item
{\tt TrafficControlDataSubmitter}
 submits data collected by ConnectionMonitor to a persistent component.
		
\item
{\tt TrafficControlDataCollector}
collects data submitted by {\tt TrafficControlData} and  creates the list of appropriate events.
	
\item {\tt TrafficControlExecutor}
performs the configuration of Traffic Control actions on devices  according to the  list of events collected by
{\tt TrafficControlDataColletor}.
\end{itemize}



Because the amount of persistent information and control information to support the provision of QoS is huge,
a Best of Practice approach (to ensure a lean operation) is to define an operation policy.
For example, a limited number of container classes (defined based on data rates) should be supported for containers,
or containers are allocated in each machine based on the number of cores and data rates (i.e., preplanned container classes
based on cores and data rates).

\begin{figure}[hbt]
		\centering
		\includegraphics[scale=0.35]{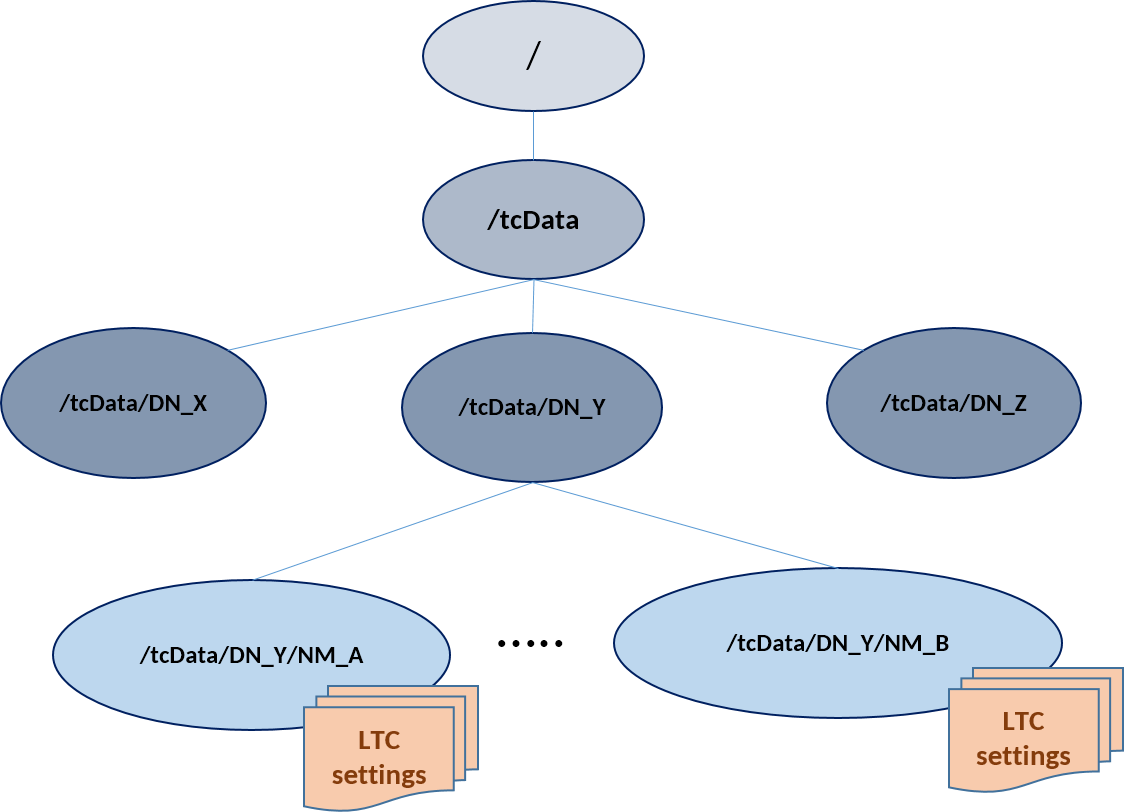}
		\caption{Data structure for storing LTC settings in ZooKeeper}
		 \label{fig:zk_storage}
\end{figure}


% ...
\subsection{Keeping the information of containers's pipes with ZooKeeper}

The maintenance of traffic control parameters and changes of the traffic control parameters
requires a persistent data structure that can be established with the use of
 Apache ZooKeeper~\cite{ZooKeeper}.
The ability to provide high availability and high performant service in distributed systems (i.e.,
to handle partial failures, to support loosely couple interactions) gives the rationale behind
the choice of ZooKeeper.

The following features of ZooKeeper are taken into account in our design:

\begin{itemize}
\item Reading/writing data of zNode is atomic, appending is not possible in ZooKeeper.
\item A zNode can be either persistent or ephemeral. A persistent zNode can be only deleted manually. An ephemeral zNode, in contrast, will be deleted if the client that created it crashes or simply closes its connection to ZooKeeper.
\item ZooKeeper deals with changes using watches. With watches, a client registers its request to receive a one-time notification of a change to a given zNode. 
\end{itemize}

In this implementation ZooKeeper acts as a persistent layer for storing and delivering LTC settings between nodes.
Note that only one ZooKeeper Server is needed for the operation (of course, additional ZooKeeper servers can be operated to increase the
reliability).
 The data structure is illustrated in Figure~\ref{fig:zk_storage}. All related traffic control data will be stored under the \textit{/tcData} root zNode. Each Datanode will register itself with ZooKeeper server by creating $/tcData/DN_{ID\_OF\_DN}$ where $ID\_OF\_DN$ is its identification, it can be the hostname or the IP address of the Datanode.

In each NodeManager, {\tt ConnectionMonitor} monitors the traffic connections and constructs the content of LTC settings and pass them to {\tt TCDataSubmitter}. Then {\tt TCDataSubmitter} notify new demands to Datanodes by creating new $/tcData/DN_{ID\_OF\_DN}/NM_{ID\_OF\_NM}$ zNode with LTC data settings (if this zNode is not existed) under the zNode node of the corresponding Datanode or replacing the data of this zNode with the new one.

In order to get new LTC settings, {\tt TCDataCollector} uses watches to collect data related to its Datanode. The first watch is put its proper $/tcData/DN_{ID\_OF\_DN}$ zNode for tracking the creation/deletion of child zNodes (e.g. whether we have new demands from new NodeManager node). A new watch will be put on each new child zNode, so we can get notifications about new demands from the corresponding NodeManager. When changes are detected, the LTC settings data will be processes by {\tt TCDataCollector}. It retrieves modifications by comparing with previous one and pass them to {\tt TCRuleExecutor} to set the LTC table.


\section{A Proof-of-concept}
\label{sec:proof}

Our proposal has been implemented within the YARN framework. The source codes of software components can be obtained from~\cite{CNTIC}.
Using machines with Intel Core i5-4670 CPU, 16GB DDR3 1600 MHz RAM memory and
1TB hard disks, we have created a small Hadoop cluster to demonstrate the capability of our building block software components.

In our testbed, we
provide an illustration to control a data rate for
the following class of applications:
\begin{itemize}
\item a job consists of multiple tasks,
\item the execution of a job can be divided into  several phases,
\item the majority of tasks should be executed within a specific phase, and  few tasks
span several phases,
\item tasks belonging to one specific phase can be executed in parallel,
each of them require one container,
\item there are HDFS intensive tasks. It is reasonable to require
that those tasks process data as  streams of bytes.
HDFS intensive tasks that are simultaneously executed
require different HDFS data blocks.
\end{itemize}

\begin{figure}[hbt]
		\centering
		\includegraphics[scale=1]{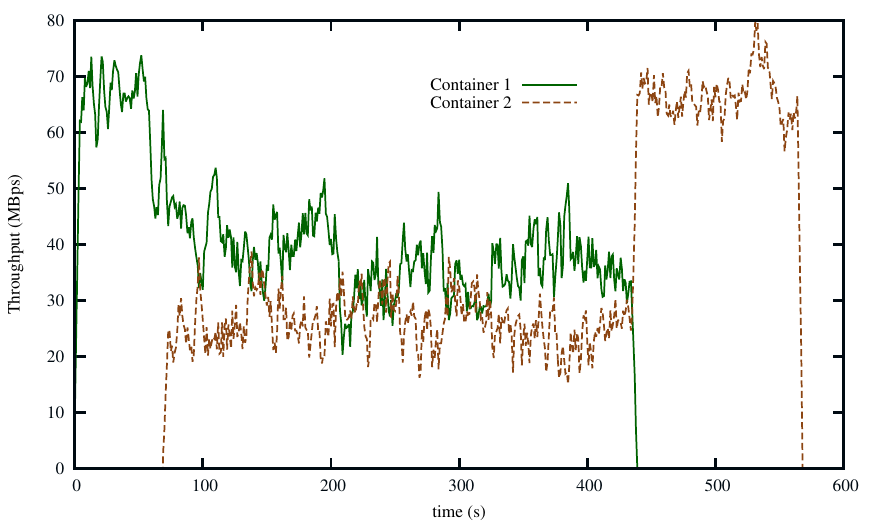}
		\caption{TCP throughput of pipes without LTC}
		 \label{fig:throughput_40mbps_p3p1_2j}
\end{figure}

\begin{figure}[hbt]
		\centering
		\includegraphics[scale=1]{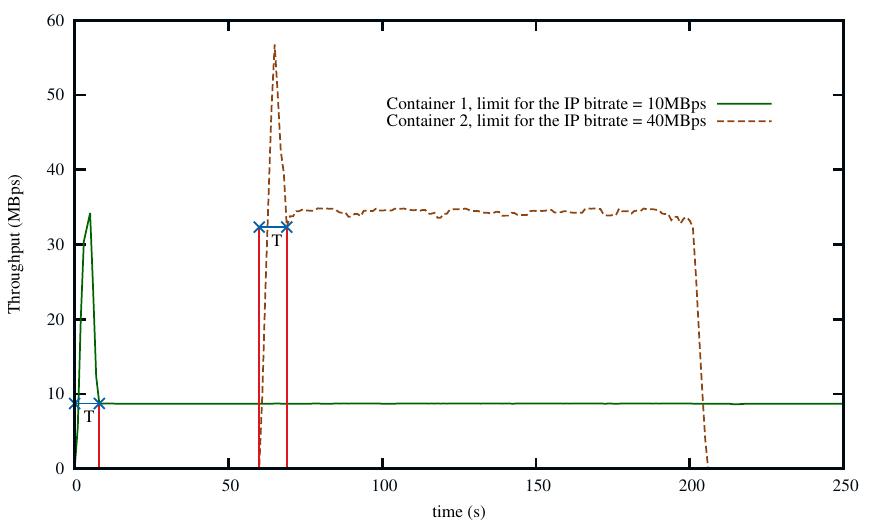}
		\caption{TCP throughput of pipes controlled by LTC}
		 \label{fig:throughput_2j_tc}
\end{figure}


Figures~\ref{fig:throughput_40mbps_p3p1_2j} and~\ref{fig:throughput_2j_tc} depict the TCP throughputs of two containers
that  read data blocks from the same DataNode. Note that the limiting rate setting in LTC is for the IP layer.
It is observed that container 1 that reads  data blocks before container 2  gets a higher throughput if no LTC is applied.
Note that disk I/O bottlenecks may happen due to certain conditions. E.g., one cause of  disk I/O bottlenecks is
the consequence of the large amount of disk blocks sequentially read by applications from an HDFS storage. In such a case,
applications that read a huge volume of data blocks from disks (I/Os per second and the amount of bytes per I/O are high) may greedily seize the whole disk I/O capacity.
The application of LTC
practically eliminates the disadvantage of the later born data pipes and
can control
the throughput of pipes in the comparison with the original YARN (where is no any limit for
pipes). Note that Figure~\ref{fig:throughput_2j_tc} also shows the greedy nature related to the I/O activity:
applications with the sequential reads of data blocks tend to capture all available I/O capacity during the uncontrolled period  if there is a  room to increase its I/O activity. From the perspective of resource management and
containers that read a huge volume of data,
 the uncontrolled period of pipes (i.e., the delay denoted by $T$ in Figure~\ref{fig:throughput_2j_tc}
from the start of containers until the LTC has the impact on the data rate
of containers) is negligible. As we mentioned earlier, the information about the
establishment of data pipes should be a part of the QoS negotiation process to eliminate
the uncontrolled period, which will be done in our future work.



\section{Conclusion}
\label{sec:con}

We have presented an approach to take into account the HDFS feature to control the throughput of data pipes between
applications and HDFS DataNodes in the YARN framework. Some basic building block functions have been implemented to exploit the property
(data pipes between DataNodes and containers)
of the HDFS architecture.
It has been shown through measurement results that the throughput of applications (jobs, containers) can be controlled within YARN
using our building block components.

At present we are working on a prototype to extend YARN. The prototype will include
the specification of I/O rates by applications,
the admission control procedure, and scheduling and management policy along with building block software components described in this paper.







\bibliographystyle{IEEEtran}
\bibliography{hadoop}
%


\end{document}